\documentclass{jps-cp-PATCHED}

\usepackage[numbers,square,sort&compress]{natbib}
\usepackage{intruder-14c-aris26}

\title{Intruder structure, deformation, and $E2$ strengths in $\isotope[14]{C}$ from an \textit{ab initio} perspective}

\author{Mark A.~\textsc{Caprio}\textsuperscript{1}}

\inst{\textsuperscript{\textup{1}}Department of Physics and Astronomy, University of Notre Dame, Notre Dame, Indiana 46556-5670, USA}

\recdate{\today}

\abst{The semimagic nucleus $\isotope[14]{C}$ lies just above the $N=8$
island of inversion,
raising the possibility of low-lying intruder states and associated deformation.
Through \textit{ab initio} no-core configuration interaction calculations, we
shed light on the role of intruder structure, quadrupole deformation, and
Elliott $\grpsu{3}$ symmetry in $\isotope[14]{C}$.
\relax
The results also highlight the influence of mixing between normal and intruder
states on the strengths of the $E2$ transitions from the first two $2^+$ states.
\relax
\relax
 }

\kword{\textit{ab initio} nuclear structure, intruder states, deformation, Elliott $\grpsu{3}$ symmetry}

\begin{document}
\maketitle

\nocite{pritychenko2016:e2-systematics,npa1991:013-015}  

\begin{figure}[b]
  \centering
  \begin{minipage}{1.0\hsize}
  \begin{minipage}{0.3\hsize}
    \includegraphics[width=\hsize]{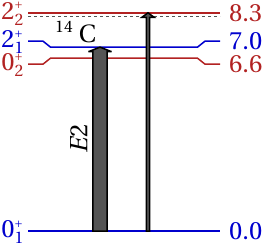}
  \end{minipage}
  \hfill
    \begin{minipage}{0.65\hsize}
      \includegraphics[width=\hsize]{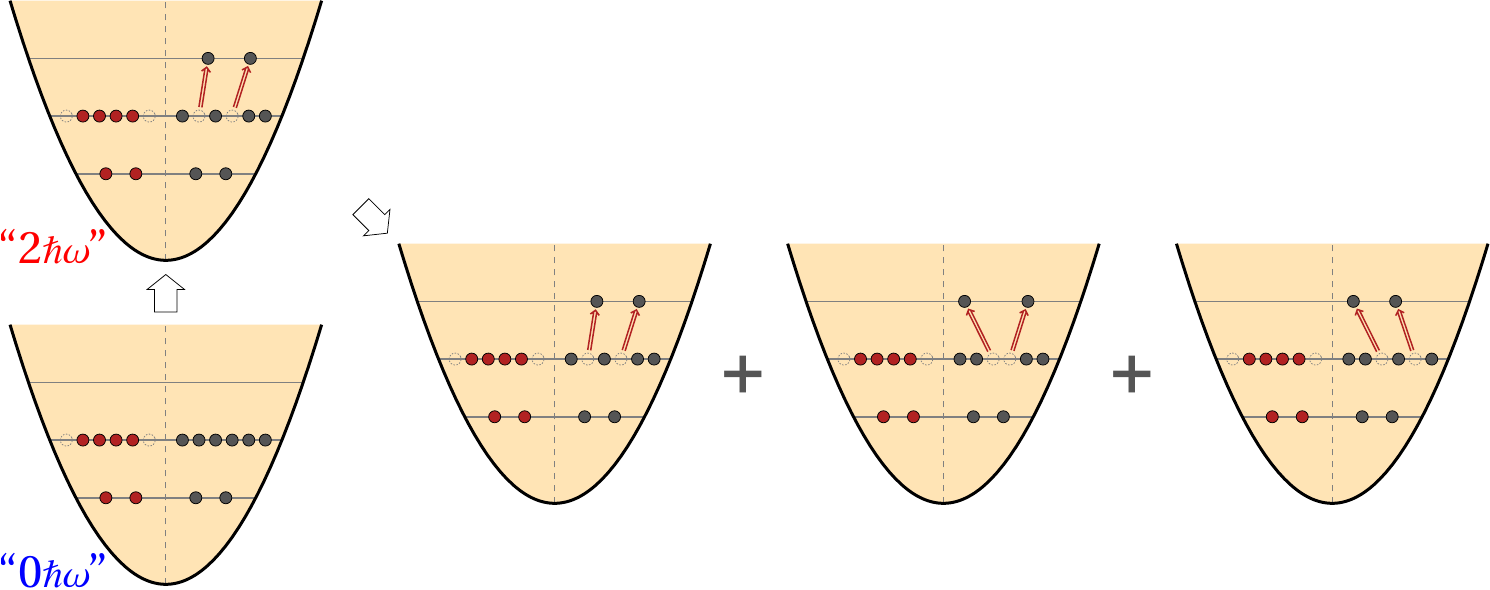}
    \end{minipage}
  \end{minipage}
  \caption{(Left)~Experimental lowest positive-parity levels of
    $\isotope[14]{C}$, with known $E2$ strengths
    indicated~\cite{npa1991:013-015}.  The neutron breakup threshold (at
    $\approx 8.2\,\MeV$) is indicated (dashed line). The suggested
    identification of the lower state of each angular momentum as normal (blue)
    and the higher as intruder (red) is indicated by the
    coloration. (Right)~Schematic diagram showing $0\hw$ and $2\hw$ shell model
    configurations for $\isotope[14]{C}$, and then a collective $2\hw$ state
    (obtained as a superposition of many two-particle, two-hole configurations) lowered due to
    increased correlation (or deformation) energy. }
  \label{fig:e2-scheme-config-diagram}
\end{figure}

\section{Introduction}
\label{sec:intro}

The excitation spectrum of $\isotope[14]{C}$ is unusual, with this nucleus
having the highest known first $2^+$ excitation energy of any even-even nucleus
other than $\isotope[4]{He}$~\cite{pritychenko2016:e2-systematics}.  The first
positive parity excited state is a $0^+$ state, followed by two relatively
closely spaced $2^+$ states, as shown in
Fig.~\ref{fig:e2-scheme-config-diagram}~(left).

There is first the question of the nature of these states, as being either
\textit{normal} or \textit{intruder}.  Two proton holes in the $p$ shell yield
two $0^+$ states and two $2^+$ states, and a $1^+$ state, as well.  However,
$\isotope[14]{C}$ is semimagic, lying on the neutron $N=8$ shell closure, just
above the $N=8$ ``island of
inversion''~\cite{sorlin2008:magic-numbers,heyde2011:shape-coexistence,nowacki2021:neutron-rich}
on the nuclear chart.  The ground states of the nearby $N=8$ nuclei
$\isotope[11]{Li}$ and $\isotope[12]{Be}$ involve two-particle, two-hole
excitations of a neutron pair to the $sd$ shell.  That is, they are intruder in
nature.  To excite two neutrons to the next major shell, as shown schematically
at left in Fig.~\ref{fig:e2-scheme-config-diagram}~(right), notionally requires
$2\hw$ in energy (where $\hw$ is the oscillator energy), which in the present
case would be $\approx 34\,\MeV$.  When a state involving such structure appears
instead in the low-lying spectrum, it is said to be an \textit{intruder} state.

Then, if there \textit{are} intruder states, there is the question of what is responsible for lowering
their energy so far.  There are at least two competing
mechanisms~\cite{nowacki2021:neutron-rich}.  In one, the
monopole interaction lowers the effective single-particle energy of a neutron
orbital from the $sd$ shell, namely $1s_{1/2}$, thereby reducing the energy
required to excite a neutron pair into that orbital.  In the other, the
quadrupole interaction is instead responsible.  The intruder state is then a
highly \textit{deformed} collective state, built from a correlated combination
of such $2\hw$ configurations, as shown schematically at right in
Fig.~\ref{fig:e2-scheme-config-diagram}~(right), and energetically favored by
the quadrupole interaction.  Exciting a neutron pair to the next major shell
effectively increases the number of valence nucleons, which enables the intruder
state to be ``lowered in energy by the enormous gain in energy resulting from
proton-pair--neutron-pair correlations''~\cite{heyde2011:shape-coexistence}.

In the present contribution, we seek to understand the role of intruder states (and deformation) in the
low-lying spectrum of $\isotope[14]{C}$ from the viewpoint of \textit{ab initio}
no-core configuration interaction (NCCI), or no-core shell model
(NCSM)~\cite{navratil2000:12c-ncsm}, calculations.  Intruder states present a
challenge to such calculations, with energies that are slow to converge with
increasing model space.  However, the use of internucleon interactions which
have been ``softened'', through similarity renormalization
group~\cite{bogner2007:srg-nucleon} transformations, enhances convergence.
Recent NCCI calculations for the lighter $N=8$ island of inversion nuclei
$\isotope[11]{Li}$~\cite{johnson2026:intruder-11li-29f} and
$\isotope[12]{Be}$~\cite{mccoy2024:12be-shape}, and nearby
$\isotope[10]{Be}$~\cite{caprio2019:bebands-sdanca19,caprio2022:10be-shape-sdanca21}
and $\isotope[11]{Be}$~\cite{caprio2020:bebands}, yield intruder states at
realistic energies.  These calculations indicate that the intruder states are
indeed deformed, in fact exhibiting the
maximum deformation possible within the $2\hw$
space~\cite{caprio2022:10be-shape-sdanca21,mccoy2024:12be-shape,johnson2026:intruder-11li-29f}.

From \textit{ab initio} NCCI calculations, we shed light on the role of intruder
states and, through Elliott's $\grpsu{3}$
symmetry~\cite{elliott1958:su3-part1,harvey1968:su3-shell-ARIS26}, deformation in the low-lying
positive-parity spectrum of $\isotope[14]{C}$ (Sec.~\ref{sec:spectrum}).  Then,
we turn to the $E2$ strengths between the ground state and both $2^+$
states, which are known from electron
scattering~\cite{crannell1972:14c-escatt-icnss-FOR-intruder-14c,fn-crannell} [arrows in
  Fig.~\ref{fig:e2-scheme-config-diagram}~(left)], and highlight their relation
to mixing between normal and intruder $2^+$ states (Sec.~\ref{sec:trans}).

\begin{figure}[tbp]
  \centering
  \begin{minipage}{1.0\hsize}
    \includegraphics[width=0.435\hsize]{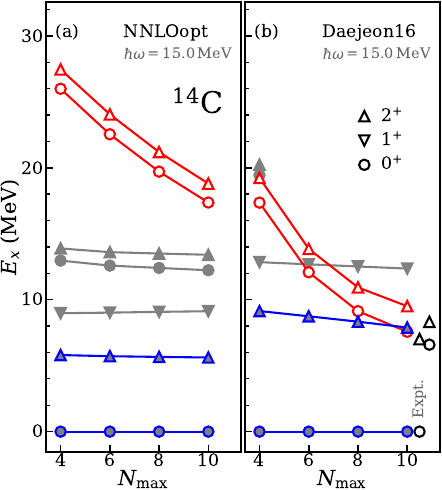}
    \hfill
    \includegraphics[width=0.26\hsize]{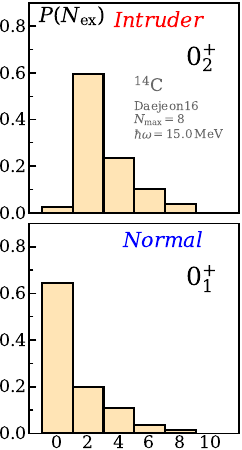}
    \hfill
    \includegraphics[width=0.26\hsize]{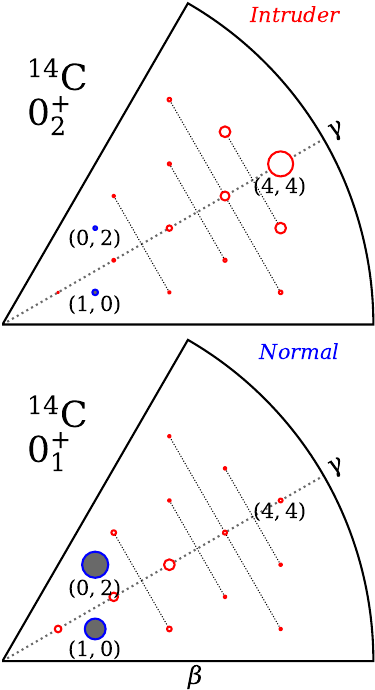}
  \end{minipage}
   \caption{(Left)~Excitation energies for selected states of $\isotope[14]{C}$,
     as calculated for the (a)~\nnloopt{} and (b)~Daejeon16 interactions.  The
     (normal) $0^+_1$ and $2^+_1$ states (blue, shaded symbols) and the first
     intruder $0^+$ and $2^+$ states (red, open symbols) are shown, along with
     the remaining normal states (gray symbols).  Results are shown as functions
     of $\Nmax$, for fixed $\hw=15\,\MeV$.  Experimental
     energies~\cite{npa1991:013-015} are shown for comparison (open symbols, at
     far right).  (Center)~Decompositions by $\Nex$ for the first two $0^+$
     states of $\isotope[14]{C}$.  Calculated for the Daejeon16 interaction, for
     $\Nmax=8$ and $\hw=15\,\MeV$.  (Right)~Decompositions jointly by $\Nex$ and
     $\grpsu{3}$ for these same two states.  The $\grpsu{3}$ irreps are arranged
     by deformation(see text), with contributions from the $0\hw$ (blue, shaded
     circles) and $2\hw$ (red, open circles) spaces shown.  Contributions from
     irreps which cannot be distinguished in the present decomposition (dotted
     lines), due to degenerate Casimir operator eigenvalues, have (arbitrarily)
     been distributed equally between these.  }
  \label{fig:spectrum}
\end{figure}

\section{Excitation spectrum and $\grpsu{3}$ structure}
\label{sec:spectrum}

Calculated excitation energies for the first normal (blue, shaded symbols) and
intruder (red, open symbols) $0^+$ and $2^+$ states in $\isotope[14]{C}$ are
shown in Fig.~\ref{fig:spectrum}~(left), from NCCI calculations 
based on either the unsoftened \nnloopt{} interaction
[Fig.~\ref{fig:spectrum}(a)] or the much softer
Daejeon16~\cite{shirokov2016:nn-daejeon16} interaction
[Fig.~\ref{fig:spectrum}(b)].  
In NCCI calculations, the many-body wave function is obtained in a basis of
$\Nex \hw$ (\textit{i.e.}, $0\hw$, $2\hw$, $\ldots$) harmonic oscillator
configurations [Fig.~\ref{fig:e2-scheme-config-diagram}~(right)], truncated to finite $\Nex
\leq \Nmax$.  Results converge towards those which would be obtained in the
full, untruncated many-body space as $\Nmax\rightarrow\infty$.  Energies in
Fig.~\ref{fig:spectrum}~(left) are shown as a function of $\Nmax$ (at a fixed
oscillator radial length scale given by $\hw=15\,\MeV$).  
Results were obtained using
\texttt{MFDn}~\cite{maris2010:ncsm-mfdn-iccs10,shao2018:ncci-preconditioned}.

Our classification of the calculated states as predominantly ``normal'' or
``intruder'', for purposes of the present discussion, is based simply
on the relative contribution of harmonic oscillator shell model configurations
of different $\Nex$ to the NCCI wave functions, illustrated for the first two
$0^+$ states in Fig.~\ref{fig:spectrum}~(center).  The first $0^+$ state
[Fig.~\ref{fig:spectrum}~(center, bottom)] obtains its largest norm, or probability,
contribution from $0\hw$ configurations, albeit with a tail of contributions from higher
$\Nex$, and is thus manifestly normal, while the second $0^+$ state in this
calculation [Fig.~\ref{fig:spectrum}~(center, top)] is largely orthogonal to the $0\hw$
configurations, instead taking its largest contribution from $2\hw$ configurations,
and is thus an intruder.

The calculated excitation energies for the normal states are nearly independent
of $\Nmax$, for either interaction, but convergence for the intruder states is
quite different.  With the \nnloopt{} interaction
[Fig.~\ref{fig:spectrum}(a)], the intruder energies start near their notional
mean-field energy of $2\hw$, at low $\Nmax$.  The energies decrease steadily
with increasing $\Nmax$, but remain far from converged, and far above the
experimental energies (open symbols, at far right) even for the largest
calculation shown ($\Nmax=10$).  However, in the calculations with the softer
Daejeon16 interaction [Fig.~\ref{fig:spectrum}(b)], the intruder excitation
energies start lower, show much greater progress towards convergence, and lie
within $\approx2\,\MeV$ of experiment by $\Nmax=10$, at which point the
intruder $0^+_2$ state has descended below the normal $2^+_1$ state, as in
experiment.

The spacing between the intruder $0^+$ and $2^+$ states stays roughly constant
with $\Nmax$, even as these states descend rapidly with respect to the normal
states, and, for either interaction, this spacing is roughly consistent with
that between the experimental $0^+_2$ and $2^+_2$ states.  It thus seems natural
to make the identification of the lower state of each angular momentum as normal
(blue) and the higher as intruder (red), as suggested in
Fig.~\ref{fig:e2-scheme-config-diagram}~(left), subject to the realization that
the $2^+$ states approach each other closely, so that normal and intruder
configurations may (and, in fact, do) mix significantly.

Elliott's $\grpsu{3}$ symmetry~\cite{elliott1958:su3-part1,harvey1968:su3-shell-ARIS26}
for the many-body problem provides a microscopic mechanism for rotation and
quadrupole deformation within the nuclear shell model.  The group generators,
consisting of the orbital angular momentum operator and the quadrupole operator
(restricted to a single shell), conserve the total number of oscillator quanta.
The $0\hw$, $2\hw$, \textit{etc.}, spaces may therefore be decomposed separately
into irreducible representations (irreps) of $\grpsu{3}$, each identified by
$\grpsu{3}$ quantum numbers $(\lambda,\mu)$.  Each irrep may, approximately, be associated with a rotational intrinsic state
with definite deformation, characterized by specific values for the Bohr
quadrupole deformation parameters $\beta$ and $\gamma$~\cite{castanos1988:su3-shape}.

Elliott's $\grpsu{3}$
model suggests that the quadrupole component of the nuclear interaction
favors the most deformed, or ``leading''~\cite{harvey1968:su3-shell-ARIS26},
$\grpsu{3}$ irrep within the shell model space.  While this model
is traditionally associated with the valence shell ($0\hw$), for intruder
states we apply this same idea to the shell model space in which the intruder
lives, in the present case, the $2\hw$ space (see
Refs.~\cite{rowe2006:coexistence-shell-u3,nowacki2021:neutron-rich}
for refinements).  We expect the leading irrep in this space to be energetically
most favored.  In the $0\hw$ space of $\isotope[14]{C}$, only the weakly deformed $(1,0)$ and $(0,2)$ irreps of
$\grpsu{3}$ are obtained.  Of the many more irreps found in the $2\hw$ space of
$\isotope[14]{C}$, the leading irrep $(4,4)$, may be expected to yield the
lowest intruder states.  This irrep has not only maximal $\beta$ deformation,
but also maximal triaxiality ($\gamma=\pi/6$).

Turning to the actual calculated NCCI calculated wave functions, these may be
decomposed into components by $\Nex (\lambda,\mu)$ via the Lanczos ``trick'',
which allows us to project the wave function into eigenspaces of the $\grpsu{3}$
quadratic Casimir
operator~\cite{gueorguiev2000:fp-su3-breaking,caprio2022:10be-shape-sdanca21,mccoy2024:12be-shape},
requiring only the machinery of a standard NCCI code.  The result is shown for
the calculated $0^+_1$ and $0^+_2$ states in Fig.~\ref{fig:spectrum}~(right),
where contributions from $0\hw$ (blue, shaded circles) and $2\hw$ (red, open
circles) irreps are arranged by quadrupole deformation in a standard polar plot
(\textit{i.e.}, with $\beta$ and $\gamma$ as radial and angular variables,
respectively).

The $0\hw$ component of the calculated ground state
[Fig.~\ref{fig:spectrum}~(right, bottom)] is largely just the naive
$jj$-coupled shell model (proton $p_{1/2}^{-2}$) state, and does not have
well-defined $\grpsu{3}$ symmetry.  However, the intruder state
[Fig.~\ref{fig:spectrum}~(right, top)] receives by far its largest
contribution from the leading $(4,4)$ irrep of the $2\hw$ space.  Thus, once
again, the intruder appears to be deformation-driven, having the maximum
possible deformation in the $2\hw$
space~\cite{caprio2022:10be-shape-sdanca21,mccoy2024:12be-shape,johnson2026:intruder-11li-29f}.
The decompositions of the first and second $2^+$ states show a similar pattern, to that for the $0^+$ states,
except that the decompositions begin to reflect strong mixing of the normal and
intruder states at higher $\Nmax$, as their energies approach.

\begin{figure}
  \centering
  \includegraphics[width=0.85\hsize]{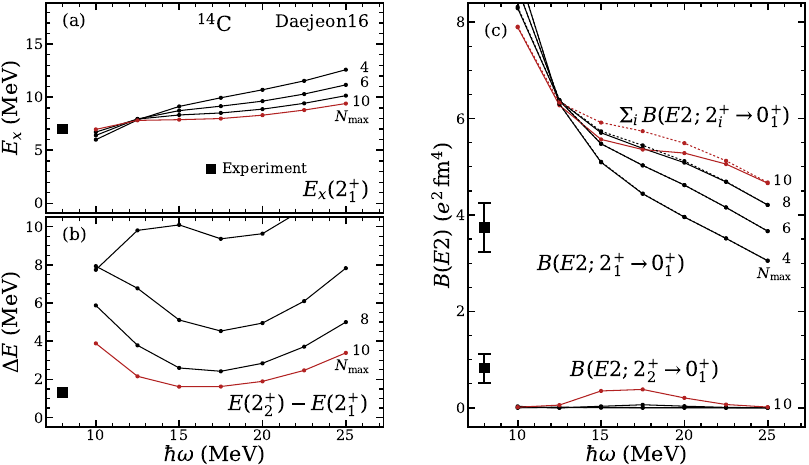}
  \caption{Calculated observables for $\isotope[14]{C}$: (a)~the $2^+_1$
    excitation energy, (b)~the energy separation of the $2^+_1$ and $2^+_2$
    levels, (c)~$B(E2;2^+_1\rightarrow0^+_1)$ and $B(E2;2^+_2\rightarrow0^+_1)$
    (solid curves, as labeled), together with the summed strength (dotted
    curve).  Values are
    shown as functions of basis parameter $\hw$ at successive $\Nmax$
    truncations (as indicated). The bare $E2$ operator is used in the calculations. Experimental results are
    provided for comparison (squares)~\cite{npa1991:013-015,crannell1972:14c-escatt-icnss-FOR-intruder-14c,fn-crannell}.
  }
  \label{fig:be2-norm-rp-scan-14c}
\end{figure}

\section{$E2$ strengths}
\label{sec:trans}

Electromagnetic transitions between normal and intruder states are typically
expected to be vanishing, and any transition strengths that are found are
indicative of mixing~\cite{heyde2011:shape-coexistence}. We now consider the
calculated strengths of the $E2$ transitions between the ground state and first
two $2^+$ states [recall Fig.~\ref{fig:e2-scheme-config-diagram}~(left)] in
Fig.~\ref{fig:be2-norm-rp-scan-14c}.  As two-state
mixing~\cite{casten2000:ns-SIMP} is crucially sensitive to the energy
difference between the states, or the ``energy
denominator'' of perturbation theory, we first examine the excitation energies.
The excitation energy of the $2^+_1$ state is now shown as a function of both
$\Nmax$ and $\hw$ [Fig.~\ref{fig:be2-norm-rp-scan-14c}(a)].  The difference in
$2^+$ energies [Fig.~\ref{fig:be2-norm-rp-scan-14c}(b)], already noted above to
decrease with increasing $\Nmax$ [Fig.~\ref{fig:spectrum}(a)], is smallest in
the range $\hw\approx15\text{--}20\,\MeV$, which is thus where we might expect
the most mixing.

The calculated $B(E2;2^+_1\rightarrow0^+_1)$ and $B(E2;2^+_2\rightarrow0^+_1)$
strengths [Fig.~\ref{fig:be2-norm-rp-scan-14c}(c)] behave at low $\Nmax$ as we
might expect without mixing: a typical $E2$ convergence
pattern (\textit{e.g.}, Ref.~\cite{caprio2020:bebands}) for the ``normal''
$2^+_1\rightarrow0^+_1$ transition, with the curves for successive $\Nmax$
gradually flattening and getting more closely spaced, while the ``intruder''
$2^+_2\rightarrow0^+_1$ transition is vanishing.  However, at higher $\Nmax$ a
``dip'' emerges in the $2^+_1\rightarrow0^+_1$ strength, and a ``bump'' emerges
in the $2^+_2\rightarrow0^+_1$ strength, aligned with the $\hw$ range in which
the energy denominator for mixing [Fig.~\ref{fig:be2-norm-rp-scan-14c}(b)] is
smallest.  Thus, it appears the ``normal'' $E2$ strength is being fragmented by
mixing~\cite{casten2000:ns-SIMP}, with a portion going to the ``intruder'' state
(we expect the total strength to be conserved).  If we take the summed $B(E2)$ [dotted line
  in Fig.~\ref{fig:be2-norm-rp-scan-14c}(c)], the ``bump'' fills in the ``dip'',
and a typical $E2$ convergence pattern is restored.

The relative strengths of the two transitions is thus a measure of the amount of
mixing.  From experiment [squares in Fig.~\ref{fig:be2-norm-rp-scan-14c}(c)],
this ratio is $\approx4.6$ to $1$, albeit with large uncertainties, in favor of
the $2^+_1$ transition.  In the NCCI calculations, the mixing is still highly
dependent upon basis parameters (through the energy denominator).

However, the summed strength, which may be taken to measure the total ``normal''
$2^+\rightarrow 0^+$ strength before fragmentation, can be compared directly
with the summed strength of $4.5(6)\,e^2\fm^4$ from experiment.  The calculated
strength, although still increasing with $\Nmax$, appears already to be above
the summed experimental values.  A similar tension with experiment is perhaps
already hinted in earlier NCCI
results~\cite{forssen2013:c-2plus-ncsm-extrapolation} with a different
interaction.  While new, more precise, measurements of the $E2$ strengths could
help clarify the situation, the neutron-unbound nature of the second $2^+$ state
(a narrow resonance, with a width of just a few
$\keV$~\cite{npa1991:013-015}) can present experimental challenges for,
\textit{e.g.}, Coulomb excitation measurements.

\section{Conclusions}
\label{sec:concl}

In \textit{ab initio} NCCI calculations for $\isotope[14]{C}$, intruder states
are found to feature prominently in the low-lying spectrum, suggesting the
second $0^+$ and $2^+$ states are intruders, from the maximally deformed
$\grpsu{3}$ irrep in the $2\hw$ space.  With normal and intruder
$2^+$ states so close in energy, the relative strengths of the $E2$ transitions
are sensitive to mixing, but the predicted total
strength exceeds experiment~\cite{crannell1972:14c-escatt-icnss-FOR-intruder-14c}.
 
\section*{Acknowledgements}
We thank Carlotta Porzio, Heather L.~Crawford, Rod M.~Clark, Paul Fallon,
Augusto O.~Machiavelli, and Anna E.~McCoy for discussions, and Peter Gysbers,
Calvin W.~Johnson, and Shwetha L.~Vittal for comments on the manuscript. This
material is based upon work supported by the U.S.~Department of Energy, Office
of Science, under Award No.~DE-FG02-95ER40934.  This research used resources of
the National Energy Research Scientific Computing Center (NERSC), a DOE Office
of Science User Facility supported by the U.S.~Department of Energy, Office of
Science, under Contract No.~DE-AC02-05CH11231, using NERSC award
NP-ERCAP0023497.

\bibliographystyle{apsrevm}  \providecommand{\newblock}{}

\begin{thebibliography}{24}
\bibitem{fn-crannell}
\bibinfo{note}{Crannell \textit{et
  al.}~\cite{crannell1972:14c-escatt-icnss-FOR-intruder-14c} report
  ``$\sqrt{B(C2,0)\uparrow}$'' values (that is, for zero momentum transfer and
  in the upward direction) of $4.32(29)\,e\fm^2$ and $2.02(38)\,e\fm^2$, from
  the ground state to the two $2^+$ states, respectively, which we may rewrite
  as $B(E2;2^+_1\rightarrow0^+_1)=3.7(5)\,e^2\fm^4$ and
  $B(E2;2^+_2\rightarrow0^+_1)=0.8(3)\,e^2\fm^4$.}
\end{thebibliography}
\providecommand{\APSLONG}{}
\providecommand{\ELSEVIER}{}

\end{document}